\documentclass[a4paper,11pt]{article}
\usepackage{amsmath}
\usepackage{graphicx}
\usepackage{amsfonts}
\usepackage{amssymb}
     \newtheorem{theorem}{Theorem}[section]

     \newenvironment{remark}[1][Remark]{\begin{trivlist}
     \item[\hskip \labelsep {\bfseries #1}]}{\end{trivlist}}
     \newcommand{\qed}{\nobreak \ifvmode \relax \else
           \ifdim\lastskip<1.5em \hskip-\lastskip
           \hskip1.5em plus0em minus0.5em \fi \nobreak
           \vrule height0.75em width0.5em depth0.25em\fi}
\numberwithin{equation}{section}

\begin{document}

\title{{\textbf{On the WDVV equations in five-dimensional gauge theories}}}
\date{}
\author{L.K. Hoevenaars, R. Martini}
\maketitle

\begin{abstract}
It is well-known that the perturbative prepotentials of four-dimensional ${\cal N}=2$
supersymmetric Yang-Mills theories satisfy the generalized WDVV equations, regardless
of the gauge group. In this paper we study perturbative prepotentials of the five-dimensional
theories for some classical gauge groups and determine whether or not they satisfy the WDVV system.
\end{abstract}

\section{Introduction}

The original WDVV equations were put forward by Witten
\cite{WITT:1991}
and Dijkgraaf, E. Verlinde and H. Verlinde
\cite{DIJK-VERL-VERL:1991}
in the context of 2-dimensional topological conformal
field theory. They form the following system of third
order nonlinear partial differential equations for
a function $F$ of $N$ variables
\begin{eqnarray}
\label{original WDVV}
F_i F_1^{-1}F_m = F_m F_1^{-1}F_i \qquad \qquad i,m=1,...,N
\end{eqnarray}
where $F_i$ is the matrix
\begin{eqnarray}
\label{matrix third order derivatives}
\left( F_i \right)_{jk} = \frac{\partial^3 F(a_1,...,a_N)}{\partial a_i 
\partial a_j \partial a_k}
\end{eqnarray}
Moreover, one requires that $F_1$ is a constant and invertible matrix.

The generalized WDVV equations however are given by the following system
\begin{eqnarray}
\label{WDVV}
F_i F_k^{-1}F_m = F_m F_k^{-1}F_i \qquad \qquad i,k,m=1,...,N
\end{eqnarray}
for any $k$, and there are no further requirements with respect to
a special coordinate. It is not difficult to show (see e.g.
\cite{MARS-MIRO-MORO:1997})
that this system can be written equivalently in a form which
is more convenient for our purposes
\begin{eqnarray}
\label{generalized WDVV}
F_i B^{-1}F_m = F_m B^{-1}F_i \qquad \qquad i,m=1,...,N
\end{eqnarray}
for a linear combination $B$ of the matrices $F_k$, possibly with $a_i$ 
dependent coefficients. In fact, (\ref{generalized WDVV}) holds
for all linear combinations $B$ simultaneously, provided their inverse exists.
The original equations (\ref{original WDVV}) are therefore indeed a special 
case of (\ref{generalized WDVV}), which explains the terminology.

The generalized WDVV system was proven to hold for prepotentials of
certain four-dimensional ${\cal N}=2$ supersymmetric Yang-Mills theories
by Marshakov, Mironov and Morozov
\cite{MARS-MIRO-MORO:1996},\cite{MARS-MIRO-MORO:1997},\cite{MARS-MIRO-MORO:2000}.
Such prepotentials consist of a perturbative and nonperturbative part,
and often it can be shown that the perturbative part itself satisfies
the WDVV system. In 
\cite{MARS-MIRO-MORO:2000},\cite{BRAD-MARS-MIRO-MORO:1999},\cite{MIRO:2000} and \cite{MIRO:PREPRINT}
the authors study perturbative prepotentials of five-dimensional
theories in combination with the WDVV equations, and it is our 
main goal in the present article to provide proofs of some of the statements
made there and to deduce new results for perturbative prepotentials
of the five-dimensional theory.

For sake of transparency we first give a summary of the results of
this paper in section \ref{section results}, followed by the central part
section \ref{section proofs} containing all the proofs.

\section{Summary of the results}
\label{section results}
In general we consider functions of the following type
\begin{eqnarray}
\label{general F}
F(a_1,...,a_N)
&=&
\sum_{1 \leq i<j \leq N} \biggl( \alpha_- f(a_i-a_j) +\alpha_+ f(a_i+a_j)  
\biggl) + \eta \sum_{i=1}^{N} f(a_i)
\nonumber \\
&+&
\frac{a}{6} \left(\sum_{i=1}^{N} a_i \right)^3 
+ \frac{b}{2} \left( \sum_{i=1}^{N}a_i \right)\left(\sum_{j=1}^N a_j^2 \right)
+ \frac{c}{6} \sum_{i=1}^N a_i^3
\end{eqnarray}
where we adopt the notation of
\cite{MARS-MIRO-MORO:2000}.
The function $f$ is defined by
\begin{eqnarray}
f(x)=\frac{1}{6}x^3 - \frac{1}{4}Li_3 (e^{-2x}) = \frac{1}{6}x^3 
-\frac{1}{4}\sum_{k=1}^{\infty} \frac{e^{-2kx}}{k^3}
\end{eqnarray}
and therefore
\begin{eqnarray}
f'''(x)=\coth(x)
\end{eqnarray}
The general form (\ref{general F}) is motivated by physics, see for instance
\cite{HARV-MOOR:1996},\cite{ANTO-FERR-TAYL:1996},\cite{NEKR:1998}.
In particular, the second line contains cubic terms coming from string theory,
serving as corrections to the naive field theoretic perturbative prepotentials.
These represent the most general cubic expression which is preserved by
permutations of the variables $a_1,...,a_N$.

For various combinations of the parameters we will investigate whether or
not $F$ satisfies the WDVV system (\ref{generalized WDVV}). The method used
involves making an appropriate choice for the matrix $B$, although the
results are of course independent of this particular choice.

\subsection{The simplest case}
\label{section simplest case}
The simplest set of parameters we consider is $\alpha_+=\eta=0$. These values 
do not correspond to an actual prepotential from physics, but we do find solutions to the WDVV system. Without loss of generality we can chose $\alpha_-=1$ by
scaling $a,b,c$.

We can prove the following result
\begin{theorem}
\label{theorem simple}
The function (\ref{general F}) with $\alpha_-=1$ , $\alpha_+=0$ and $\eta=0$ 
satisfies the WDVV system (\ref{generalized WDVV}) if and only if the
following relation holds
\begin{eqnarray}
Nb^3+3b^2c-ac^2+3Nb+c+N^2a=0
\end{eqnarray}
\end{theorem}
More accurately, this relation is correct in the generic case that both $Nb+c \neq 0$ and $Na+2b \neq 0$. Special cases will be discussed separately in section
\ref{section proof special values}.

\subsection{The type $A$ prepotential}
\label{section type A}
Let us now turn to a prepotential with physical background. We consider
the function
\begin{eqnarray}
\label{try F A type}
{\tilde F}(x_1,...,x_{N+1})=\sum_{1\leq i<j \leq N+1}f(x_i-x_j) +
\frac{N+1}{2} \sum_{1\leq i<j<k \leq N+1} x_ix_jx_k
\end{eqnarray}
which is of the form of the previous paragraph with parameters $a,b,c$ given by
\begin{eqnarray}
a=\frac{N+1}{2} \qquad b=-\frac{N+1}{2} \qquad c=N+1
\end{eqnarray}
The $SU(N+1)$ perturbative prepotential is obtained from ${\tilde F}$ by the 
linear \footnote{The WDVV equations are invariant under linear coordinate 
changes} change of variables
\begin{eqnarray}
a_i
&=&
x_i-x_{N+1} \qquad \qquad \qquad i=1,..,N
\nonumber \\
a_{N+1}&=&x_1+...+x_{N+1} 
\end{eqnarray}
and the substitution $a_{N+1}=0$. Concretely it is given by
\begin{eqnarray}
F(a_1,...,a_{N})
&=&
\sum_{1 \leq i<j \leq N}f(a_i-a_j) + \sum_{i=1}^{N} f(a_i)
\nonumber \\
&+&
\frac{1}{3(N+1)} \left(\sum_{i=1}^{N} a_i \right)^3 
- \frac{1}{2} \left( \sum_{i=1}^{N}a_i \right)\left(\sum_{j=1}^N a_j^2 \right)
+ \frac{N+1}{6} \sum_{i=1}^N a_i^3
\end{eqnarray}
This is of the general type (\ref{general F}) with parameters
\begin{eqnarray*}
\begin{tabular}{|c|c|c|}
\hline
$\alpha_-=1$      & $\alpha_+=0$ & $\eta =1$ \\
\hline
$a=\frac{2}{N+1}$ & $b=-1$       & $c=N+1$ \\
\hline
\end{tabular}
\end{eqnarray*}
It turns out that the sign of the correction term in (\ref{try F A type}) is irrelevant for the WDVV equations. 

We can confirm the result in
\cite{MARS-MIRO-MORO:2000}
and prove
\begin{theorem}
\label{theorem A type}
The function
\begin{eqnarray}
\label{F A type}
F(a_1,...,a_{N})
&=&
\sum_{1 \leq i<j \leq N}f(a_i-a_j) + \sum_{i=1}^{N} f(a_i)
\nonumber \\
&\pm&
\frac{1}{3(N+1)} \left(\sum_{i=1}^{N} a_i \right)^3 
\mp \frac{1}{2} \left( \sum_{i=1}^{N}a_i \right)\left(\sum_{j=1}^N a_j^2 \right)
\pm \frac{N+1}{6} \sum_{i=1}^N a_i^3
\end{eqnarray}
satisfies the WDVV system (\ref{generalized WDVV}).
\end{theorem}
\begin{remark}
We note that (\ref{F A type}) is invariant under the Weyl group of $A_N$. In fact, taking arbitrary values for $a,b,c$ this is still the case. A natural
question is therefore whether an $F$ with $\alpha_-=1$, $\alpha_+=0$ and $\eta=1$ satisfies the WDVV system for any other values of $a,b,c$. Calculations
for ranks up to five suggest that there are no other solutions. This 
should mean that the string theory corrections are {\emph{precisely}} the ones
needed to satisfy the WDVV equations.
\end{remark}

\subsection{Other classical Lie algebras}
\label{section BCD type}
Next we consider a prepotential inspired by the other classical Lie algebras.
Without correction terms, the $B,D$ prepotentials are given by $\alpha_-=1$, $\alpha_+=1$ and $\eta=1,0$ respectively.
Leaving the parameter $\eta$ unfixed, we can prove the following theorem
\begin{theorem}
\label{theorem BCD type}
The function
\begin{eqnarray}
\label{F BCD type}
F(a_1,...,a_{N}) = \sum_{1 \leq i<j \leq N} \biggl( f(a_i-a_j)+f(a_i+a_j)  \biggl) +\eta \sum_{i=1}^{N}f(a_i)
\end{eqnarray}
satisfies the WDVV equations (\ref{generalized WDVV}) if and only if $\eta=-2(N-2)$.
\end{theorem}
\begin{remark}
This solution seems to have little to do with the $B,D$ Lie algebras, since 
$\eta$ is different from $1$ and $0$.
One can think about adding correction terms to restore the Lie algebraic interpretation, but the 
Lie algebras under consideration do not possess any third order
Weyl invariant polynomials. 
This forces one to consider fourth (or higher) order correction terms
and therefore nonconstant additions to the third order derivatives of $F$, which is beyond
the scope of this article. For the sake of completion we mention that calculations for low
ranks indicate that adding third order correction terms with parameters $a,b,c$ indeed
doesn't help: the WDVV equations always force $a=b=c=0$.
\end{remark}

\subsubsection{Adding an extra variable}
\label{adding}
We can add a variable $a_{N+1}$ to the prepotential (\ref{F BCD type}) and obtain the following result
\begin{theorem}
\label{theorem adding}
The function
\begin{eqnarray}
\label{F adding}
F(a_1,...,a_{N+1})=\sum_{1 \leq i<j \leq N} \biggl( f(a_i-a_j)+f(a_i+a_j)  \biggl) +\eta \sum_{i=1}^{N}f(a_i)
+\frac{\gamma}{6} ( a_{N+1}^3+3a_{N+1}\sum_{i=1}^{N}a_i^2  )
\end{eqnarray}
satisfies the WDVV system (\ref{generalized WDVV}) if and only if
\begin{eqnarray}
\eta=-2(N-2) -\frac{\gamma^2}{2}
\end{eqnarray}
\end{theorem}
With respect to the variable $a_{N+1}$ the matrix of third order derivatives is
\begin{eqnarray}
\left(  F_{N+1} \right)_{ij}=\frac{\partial^3 F}{\partial a_{N+1}\partial a_i\partial a_j}
\end{eqnarray}
and therefore becomes a multiple of the identity. So the function (\ref{F adding}) can even be regarded
as a solution to the original WDVV equations (\ref{original WDVV}). To the best of our
knowledge such functions, both as a solution to the generalized system and to the original one, have
not been considered before.

\subsection{An additional result}
\label{section additional result}
In this section we mention a result which was obtained in the process of proving the main
results of this article. The five-dimensional prepotentials are built from a basic function
$f$ with $f'''(x)=\coth(x)$. The four-dimensional theories have a basic function with
$f'''(x)=\frac{1}{x}$. We can prove the following result, which should be compared with
theorem \ref{theorem simple}
\begin{theorem}
\label{theorem 1/x}
The function (\ref{general F}) with $\alpha_-=0$, $\alpha_+=1$, $\eta=0$ and $f'''(x)=\frac{1}{x}$ 
satisfies the WDVV equations (\ref{generalized WDVV}) if and only if
\begin{eqnarray}
Nb^3+3b^2c-ac^2=0
\end{eqnarray}
\end{theorem}
In the next sections, we will describe the proofs of theorems \ref{theorem simple}-\ref{theorem 1/x}.

\section{The proofs}
\label{section proofs}
The general idea underlying all the proofs of this section is to find an appropriate linear combination $B$ such that its inverse, appearing in (\ref{generalized WDVV}), becomes manageable. 
In the four-dimensional theory it is possible to chose $B$ to equal the Killing form of the Lie algebra
under consideration
\cite{MART-GRAG:1999}.
In our five-dimensional setting this is no longer the case. This makes the Lie algebraic structure less
transparent than in the four-dimensional case.

We start by proving theorems \ref{theorem simple} and \ref{theorem 1/x} in section \ref{section proof simple}, making use of a constant nondiagonal matrix $B$. Theorems \ref{theorem A type}, \ref{theorem BCD type} and \ref{theorem adding} on the other hand are proven by taking a diagonal (not always constant) $B$, which allows for a common general strategy for these cases, given in section \ref{section general strategy}. The individual theorems are subsequently proven in sections \ref{section proof A type},\ref{section proof BCD type} and \ref{section proof adding} respectively. 

\subsection{The simplest case}
\label{section proof simple}
We will prove theorem \ref{theorem simple}. Therefore we consider a function of the form
\begin{eqnarray}
F(a_1,...,a_N)= \sum_{1 \leq i<j \leq N}  f(a_i-a_j)
+\frac{a}{6} \left(\sum_{i=1}^{N}a_i \right)^3 + \frac{b}{2} \left( \sum_{i=1}^{N}a_i \right)\left(\sum_{j=1}^N a_j^2 \right)
+ \frac{c}{6} \sum_{i=1}^N a_i^3
\end{eqnarray}
where
\begin{eqnarray}
\label{antisymmetric}
f'''(-x)=-f'''(x)
\end{eqnarray}
Writing $\beta_{ij}=f'''(a_i-a_j)$, the third order derivatives of $F$ are
\begin{eqnarray}
F_{klm}
&=&
a+\delta_{kl}\delta_{lm} \left( \sum_{q \neq k}\beta_{kq} +3b+c \right) +\delta_{kl}(1-\delta_{km})(\beta_{mk}+b)
\nonumber \\
&+&
\delta_{km}(1-\delta_{kl})(\beta_{lk}+b)+\delta_{lm}(1-\delta_{kl})(\beta_{kl}+b)=aU_{lm}+\left(V_k \right)_{lm}
\end{eqnarray}
We take a specific linear combination $B=\sum_{j=1}^{N}F_j$ and using (\ref{antisymmetric}) we find
\begin{eqnarray}
\label{B simple}
B=(Na+2b)U+(Nb+c)I
\end{eqnarray}
Special situations occur when $Na+2b=0$ and / or $Nb+c=0$. The first results in $B$ being a multiple
of the identity and the second causes $B$ to become singular. For the moment we will work with
generic $B$ and we will come back to the special cases later. The inverse of $B$ equals up to a factor
\begin{eqnarray}
B^{-1}_{kl}=1+\delta_{kl} \left( -\frac{Nb+c}{Na+2b}-N \right)
\end{eqnarray}
For the WDVV equations to hold, we should have
\begin{eqnarray}
\left(F_i B^{-1}F_m \right)_{jn} -\left( F_mB^{-1}F_i \right)_{jn} =0
\end{eqnarray}
or equivalently
\begin{eqnarray}
\label{condition WDVV simple}
B_{ij}B_{mn}-B_{mj}B_{in}-\frac{3Nb+c+N^2a}{Na+2b}\left[ F_i,F_m \right]_{jn} =0
\end{eqnarray}
We will first calculate the commutator $[F_i,F_m]=\left[ aU+V_i,aU+V_m \right]$. We find
\begin{eqnarray}
\left( UV_m \right)_{kl} = 2b+\delta_{lm}(1-\delta_{kl})(Nb+c)
\end{eqnarray}
and since $U^T=U$ and $V_m^T=V_m$ we also know $V_m U=(UV_m)^T$. Furthermore, if we use
the identity
\begin{eqnarray}
\beta_{ij}\beta_{ik}+\beta_{ij}\beta_{kj}+\beta_{ik}\beta_{jk}=1
\end{eqnarray}
we find
\begin{eqnarray}
\left[ V_i,V_m  \right]_{jn} 
&=&
\delta_{ij}(1-\delta_{mn})(1-\delta_{in})(b^2-1)
+\delta_{mn}(1-\delta_{jm})(1-\delta_{ij})(b^2-1)
\nonumber \\
&-& \delta_{jm}(1-\delta_{mn})(1-\delta_{in})(b^2-1)
-\delta_{in}(1-\delta_{jm})(1-\delta_{ij})(b^2-1)
\nonumber \\
&+&\delta_{ij}\delta_{mn}(\beta +2(b^2-1))
-\delta_{jm}\delta_{in}(\beta+2(b^2-1))
\end{eqnarray}
and therefore
\begin{eqnarray}
\label{commutator simple}
\left[ F_i,F_m \right]_{jn}
&=&\delta_{ij}(1-\delta_{mn})(1-\delta_{in})\alpha
+\delta_{mn}(1-\delta_{jm})(1-\delta_{ij})\alpha-\delta_{jm}(1-\delta_{mn})(1-\delta_{in})\alpha
\nonumber \\
&-&
\delta_{in}(1-\delta_{jm})(1-\delta_{ij})\alpha+\delta_{ij}\delta_{mn}(\beta -2\alpha)
-\delta_{jm}\delta_{in}(\beta-2\alpha)
\end{eqnarray}
where
\begin{eqnarray}
\alpha &=& b^2-1-ac-Nab
\nonumber \\
\beta &=& N+Nb^2+2bc
\end{eqnarray}
On the other hand, we have
\begin{multline} \label{B relatie simple}
B_{ij}B_{mn}-B_{mj}B_{in}=(Na+2b)^2 \Biggl[
\delta_{ij}(1-\delta_{mn})(1-\delta_{in})\gamma
+\delta_{mn}(1-\delta_{jm})(1-\delta_{ij})\gamma
\\
-\delta_{jm}(1-\delta_{mn})(1-\delta_{in})\gamma
- \delta_{in}(1-\delta_{jm})(1-\delta_{ij})\gamma+\delta_{ij}\delta_{mn}(\delta -2\gamma)
-\delta_{jm}\delta_{in}(\delta-2\gamma)
\Biggl]
\end{multline}
where
\begin{eqnarray}
\gamma&=&\frac{Nb+c}{Na+2b}
\nonumber \\
\delta&=&\gamma^2
\end{eqnarray}
The equation (\ref{condition WDVV simple}) therefore reduces to two algebraic relations
among the parameters $a,b,c$. These relations are
\begin{eqnarray}
-\frac{3Nb+c+N^2a}{Na+2b}\alpha + (Na+2b)^2\gamma =0
\end{eqnarray}
and
\begin{eqnarray}
-\frac{3Nb+c+N^2a}{Na+2b}(2\alpha-\beta) + (Na+2b)^2(2\gamma-\delta) =0
\end{eqnarray}
which combine into only one relation
\begin{eqnarray}
\label{result simple}
Nb^3+3b^2c-ac^2+3Nb+c+N^2a=0
\end{eqnarray}
This finishes the proof of theorem \ref{theorem simple}. In the next section we will
look at the nongeneric values of the parameters $a,b,c$.

\subsubsection{Special values for $a,b,c$.}
\label{section proof special values}
Recall from (\ref{B simple}) that there are special situations for either
$Na+2b=0$ or $Nb+c=0$ or both. If $Na+2b=0$ and $Nb+c \neq 0$ then we find
that the WDVV equations hold if and only if
\begin{eqnarray}
\label{commutator}
\left[ F_i,F_m\right] =0
\end{eqnarray}
and therefore if and only if
\begin{eqnarray}
\alpha = 1 +\left( \frac{Na}{2} \right)^2-ac=0
\end{eqnarray}
and
\begin{eqnarray}
2\alpha-\beta = -(N-2)\left(1+ \left( \frac{Na}{2} \right)^2-ac \right) =0
\end{eqnarray}
Note that just substituting $b=-\frac{Na}{2}$ in (\ref{result simple}) gives
\begin{eqnarray}
(N^2a-2c) \left( 1+ \left( \frac{Na}{2} \right)^2-ac  \right)=0
\end{eqnarray}
which is only partially correct since $N^2a-2c=0$ does not yield a solution.

Furthermore, if $Nb+c=0$ and $Na+2b \neq0$, then (\ref{B simple}) shows that $B$ becomes 
singular. Experience tells us that for $N \neq 3$ there exist no solutions to the WDVV
equations without the extra requirement $b=\pm 1$. For $N=3$ there is no such
condition on $b$ and the WDVV equations are satisfied. We will now consider 
$b=1$ and $b=-1$ separately. If $b=1$ then we chose a new nonsingular $B$ equal to
\begin{eqnarray}
B=\sum_{j=1}^{N}h_jF_j = \sum_{j=1}^{N} \left(-(2+a(N-1))e^{2a_j}+a\sum_{i\neq j}e^{2a_i} \right)F_j
\end{eqnarray}
and working this out we find that $B$ equals up to a factor
\begin{eqnarray}
\left(\frac{Na}{2}+b\right) I
\end{eqnarray}
which is a nonzero multiple of the identity since $Na+2b \neq 0$. If $b=-1$ on the other hand,
we take
\begin{eqnarray}
B=\sum_{j=1}^{N}h_jF_j = \sum_{j=1}^{N} \left(\prod_{k\neq j}(2-a(N-1))e^{2a_k}+
a\sum_{k\neq j}\prod_{i\neq k}e^{2a_i}\right)F_j
\end{eqnarray}
which also leads to $B$ being a multiple of the identity. So in both cases we must solve
(\ref{commutator}) again, which leads to
\begin{eqnarray}
Na+2b=0
\end{eqnarray}
which is precisely what we excluded before.

Finally, if we take both $Na+2b=0$ and $Nb+c=0$ then all linear combinations of the $F_j$ become singular
and the WDVV equations are meaningless.

Summarizing, we conclude that if $Na+2b=0$ and $Nb+c \neq 0$ there are solutions if and only if
\begin{eqnarray}
 1+ \left( \frac{Na}{2} \right)^2-ac =0
\end{eqnarray}
and if $Nb+c=0$ and $Na+2b \neq 0$ there are solutions if and only if $N=3$ and finally if both 
$Na+2b=0$ and $Nb+c=0$ then there are no solutions at all.

This finishes the discussion of theorem \ref{theorem simple}.

\subsubsection{Taking $f'''(x)=\frac{1}{x}$ instead of $\coth(x)$}
\label{section proof 1/x}
In the previous section, we took $f'''(x)=\coth(x)$ which leads to the relation
\begin{eqnarray}
  \beta_{ij}\beta_{ik}+\beta_{ij}\beta_{kj}+\beta_{ik}\beta_{jk}=1
\end{eqnarray}
and under the condition $f'''(-x)=-f'''(x)$ this is the only differentiable solution to this relation. We could also consider instead the equation
\begin{eqnarray}
  \beta_{ij}\beta_{ik}+\beta_{ij}\beta_{kj}+\beta_{ik}\beta_{jk}=0
\end{eqnarray}
and assuming again $f'''(-x)=-f'''(x)$ we find that the only solution is $f'''(x)=\frac{1}{x}$. This is the basic function for the four-dimensional theory. Adding correction terms, we find that $[F_i,F_m]$ is of the same form as (\ref{commutator simple}) but with
\begin{eqnarray}
  \alpha &=& -b^2-ac-Nab \nonumber \\
  \beta &=& Nb^2+2bc
\end{eqnarray}
and we find precisely the same $B_{ij}B_{mn}-B_{mj}B_{in}$ as in (\ref{B relatie simple}). For $N\neq 2$ this again leads to a single relation, namely
\begin{eqnarray}
  Nb^3+3b^2c-ac^2=0
\end{eqnarray}
which is to be compared with (\ref{result simple}). This finishes the proof of theorem \ref{theorem 1/x}.

\subsection{General strategy}
\label{section general strategy}
In the previous section proofs were given of theorems \ref{theorem simple} and \ref{theorem 1/x}. To discuss the proofs of the remaining theorems we will use the following general strategy. We take the general function of (\ref{general F}) and return to the case $f'''(x)=\coth(x)$. This $F$ has third order derivatives equal to
\begin{eqnarray}
\label{third order derivatives}
  F_{klm}=a+\delta_{kl}\delta_{lm}K_k+\delta_{kl}\beta_{mk} +\delta_{km}\beta_{lk}+\delta_{lm}\beta_{kl}
\end{eqnarray}
where
\begin{eqnarray}
  K_k
&=&
\sum_{q \neq k}\beta_{kq}+\beta_k
\nonumber \\
\beta_k
&=&
\eta \coth(a_k)+(4-N)b+c
\nonumber \\
\beta_{ij}
&=&
\Biggl\{
\begin{array}{cc}
0 & \quad $if$ \quad i=j \\
\alpha_- \coth(a_i-a_j)+\alpha_+\coth(a_i+a_j)+b & \quad $if$ \quad i\neq j
\end{array}
\end{eqnarray}
Consider a linear combination $B$ of the following form
\begin{eqnarray}
\label{general B}
  B_{kl} = \delta_{kl}\frac{1}{A_k}
\end{eqnarray}
where $A_k$ depends on the specific prepotential under consideration and will be specified later. We find
\begin{eqnarray}
  \left( F_i B^{-1}F_m \right)_{jl}
&=&
\sum_{k=1}^{N}F_{ijk}A_kF_{klm} 
\nonumber \\
&=&
\delta_{im}\biggl( A_i \beta_{ji}\beta_{lm} + \delta_{ij}\beta_{lm}A_iK_i
+\delta_{il}\beta_{ji}A_iK_i + \delta_{ij}\delta_{lm} A_iK_iK_m \biggl)
\nonumber \\
&+&
\delta_{jl}(1-\delta_{lm})A_j\beta_{ij}\beta_{mj}
\nonumber \\
&+&
\delta_{jl}\delta_{lm}A_mK_l\beta_{im}
\nonumber \\
&+&
\delta_{ij}\delta_{il}A_iK_l\beta_{mi}
\nonumber \\
&+&
\delta_{il}(1-\delta_{jm})A_i\beta_{ji}\beta_{mi}
\nonumber \\
&+&
\delta_{lm}(1-\delta_{ij})\left( A_i\beta_{ji}\beta_{im}+A_j\beta_{ij}\beta_{jm} +aA_mK_m+a\sum_k A_k\beta_{km} \right)
\nonumber \\
&+&
\delta_{jm}(1-\delta_{il})A_j \beta_{ij} \beta_{lj}
\nonumber \\
&+&
\delta_{ij}(1-\delta_{lm})\left( A_l \beta_{li}\beta_{ml} + A_m\beta_{mi}\beta_{lm} aA_iK_i+a\sum_k A_k\beta_{ki}  \right)
\nonumber \\
&+&
\delta_{ij}\delta_{lm} \biggl( A_iK_i\beta_{im} +A_mK_m\beta_{mi}
+\sum_{k\neq i,m}A_k\beta_{km}\beta_{ki} +aA_mK_m 
\nonumber \\
&&\qquad \qquad \qquad \qquad
+a\sum_{k}A_k\beta_{km} +aA_iK_i +a\sum_k A_k\beta_{ki}
\biggl)
\nonumber \\
&+&
\delta_{jm}\delta_{il} \left(A_m\beta_{im}^2 + A_i \beta_{mi}^2 \right)
\nonumber \\
&+&
a^2\sum_k A_k +a(A_j\beta_{ij}+A_i\beta_{ji}) +a(A_m\beta_{lm}+A_l\beta_{ml})
\label{general expression}
\end{eqnarray}
Here it should be noted that the last line contributes to all the previous ones. For example, if $i=l,i\neq m,i\neq j,j\neq m$ then 
$\left( F_i B^{-1} F_m \right)_{jl}$ is not
\begin{eqnarray}
  A_j \beta_{ij}\beta_{mj}
\end{eqnarray}
but
\begin{eqnarray}
  A_j \beta_{ij}\beta_{mj}+ a^2\sum_k A_k +a(A_j\beta_{ij}+A_i\beta_{ji}) +a(A_m\beta_{im}+A_i\beta_{mi})
\end{eqnarray}
In order to satisfy the WDVV system we should check whether or not (\ref{general expression}) is symmetric in $i$ and $m$. For example, the first two lines automatically are preserved under the interchange of $i$ and $m$. The third and fourth lines on the other hand are mutually exchanged. The rest of condition (\ref{generalized WDVV}) is nontrivial and depends on the details of the function $F$.

\subsection{The type $A$ prepotential}
\label{section proof A type}
As mentioned in section \ref{section type A}, after taking $\alpha_-=1$, $\alpha_+=0$ and $\eta=1$ the only solutions to the WDVV equations exist for the parameters
\begin{eqnarray}
  a=\pm \frac{2}{N+1} \qquad , \qquad b=\mp 1 \qquad , \qquad c=\pm(N+1)
\end{eqnarray}
These two cases will be treated separately in the following sections.

\subsubsection{The parameters $a=-\frac{2}{N+1}$ , $b=1$ , $c=-N-1$}
\label{section proof A type parameters 1}
The third order derivatives of $F$ are given by
\begin{eqnarray}
  F_{klm}=-\frac{2}{N+1} +\delta_{kl}\delta_{lm}K_k+\delta_{kl}\beta_{mk} +\delta_{km}\beta_{lk}+\delta_{lm}\beta_{kl}
\end{eqnarray}
with
\begin{eqnarray}
  K_k
&=&
\sum_{q \neq k}\beta_{kq}+\beta_k
\nonumber \\
\beta_k
&=&
\frac{2}{1-e^{-2a_k}}-2(N-1)
\nonumber \\
\beta_{ij}
&=&
\Biggl\{
\begin{array}{cc}
0 & \quad $if$ \quad i=j \\
2\frac{e^{2a_i}}{e^{2a_i}-e^{2a_j}} & \quad $if$ \quad i\neq j
\end{array}
\end{eqnarray}
We take a specific linear combination $B=\sum_j h_j F_j$ where
\begin{eqnarray}
  h_j = e^{2a_j}+\sum_{i=1}^{N}e^{2a_i}
\end{eqnarray}
and we find up to a factor
\begin{eqnarray}
  B_{kl}=\delta_{kl}\left( \frac{1}{1-e^{-2a_k}} \right) =\delta_{kl}\frac{1}{A_k}
\end{eqnarray}
Using this information we can derive the following identities
\begin{eqnarray}
\label{type A relatie 1}
A_j\beta_{ij} +A_i\beta_{ji} &=& 2-\frac{2}{e^{2a_i}}-\frac{2}{e^{2a_j}}
\\
\label{type A relatie 2}
A_i \beta_{ij}+A_j\beta_{ji}&=&2
\\
\label{type A relatie 3}
A_i\beta_{ji}\beta_{im} + A_j\beta_{ij}\beta_{jm}-A_m\beta_{jm}\beta_{im}&=&\frac{4}{e^{2a_m}}
\end{eqnarray}
Turning to the WDVV condition we find that the first two lines of (\ref{general expression}) are preserved under the interchange of $i$ and $m$ and that the third and fourth lines become mutually exchanged. We will now study the fifth and sixth lines. Keeping in mind that the last line of (\ref{general expression}) contributes to both of these, we find that the fifth line becomes
\begin{eqnarray}
  \delta_{il}(1-\delta_{lm})\left( A_j\beta_{ij}\beta_{mj}+a^2\sum_k A_k 
+a\left(A_j\beta_{ij}+A_i\beta_{ji} \right) +a \left( A_m\beta_{lm}+A_l\beta_{ml} \right) \right)
\nonumber 
\end{eqnarray}
and the sixth becomes
\begin{eqnarray}
  \delta_{lm}(1-\delta_{ij})\left( A_i\beta_{ji}\beta_{im}+ A_j\beta_{ij}\beta_{jm}+ a^2\sum_k A_k
+a\left(A_j\beta_{ij}+A_i\beta_{ji} \right)
+a\sum_k A_k \beta_{km} + aA_mK_m
\right)
\nonumber
\end{eqnarray}
Using the definition of $K_m$ and the relations (\ref{type A relatie 1}), (\ref{type A relatie 2}) and (\ref{type A relatie 3}) we see that these are indeed exchanged under the interchange of $i$ and $m$. The seventh and eighth lines of (\ref{general expression}) are mutually exchanged for the same reasons, which leaves us with the ninth and tenth lines. The complete ninth line becomes
\begin{eqnarray}
\label{ninth line 1}
  \delta_{ij}\delta_{lm} 
&\biggl(& 
A_iK_i\beta_{im} 
+A_mK_m\beta_{mi}
+\sum_{k\neq i,m}A_k\beta_{km}\beta_{ki} 
+aA_mK_m 
+a\sum_{k}A_k\beta_{km} 
\nonumber \\
&& +aA_iK_i 
+a\sum_k A_k\beta_{ki}
+a^2\sum_k A_k
\biggl)
\end{eqnarray}
and the tenth line is
\begin{eqnarray}
\label{tenth line 1}
  \delta_{jm}\delta_{il} \left(A_m\beta_{im}^2 + A_i \beta_{mi}^2 
+2a\left( A_m\beta_{im} + A_i \beta_{mi} \right)
+ a^2\sum_k A_k
\right)
\end{eqnarray}
Using the definition of $K_m$ and working out (\ref{ninth line 1}) we find
\begin{eqnarray}
  \delta_{ij}\delta_{lm} 
&\Biggl(&
\sum_{k\neq i,m} \left( A_k\beta_{ik}\beta_{im}+A_m\beta_{mk}\beta_{mi}
+A_k\beta_{km}\beta_{ki} \right)
+A_i\beta_{im}^2
+A_m\beta_{mi}^2
+A_i\beta_i\beta_{im}
+A_m\beta_m\beta_{mi}
\nonumber \\
&+&
a\sum_{k \neq m} \left( A_m\beta_{mk}+A_k\beta_{km} \right)
+aA_m\beta_m + a\sum_{k \neq i} \left( A_i\beta_{ik}A_k\beta_{ki} \right)
+aA_i\beta_i
+a^2\sum_k A_k
\Biggl)
\end{eqnarray}
We will make use of (\ref{type A relatie 2}) and the following relations
\begin{eqnarray}
  \label{type A relatie 4}
A_i \beta_{ik}\beta_{im} + A_m\beta_{mk}\beta_{mi} + A_k\beta_{km}\beta_{ki}&=&4
\\
\label{type A relatie 5}
A_i\beta_{i}\beta_{im}+A_m\beta_m\beta_{mi} &=& 8-4N
\end{eqnarray}
and we find that the ninth line becomes
\begin{eqnarray}
  \delta_{ij}\delta_{lm} \left(
A_i\beta_{im}^2 + A_m\beta_{mi}^2 +4a(N-1) +a\left(A_m\beta_m +A_i \beta_i \right)+a^2\sum_k A_k
\right)
\end{eqnarray}
Using (\ref{type A relatie 2}) again we find that the tenth line becomes
\begin{eqnarray}
  \delta_{jm}\delta_{il} \left(
A_m\beta_{im}^2+A_i\beta_{mi}^2+4a-\frac{4a}{e^{2a_i}}-\frac{4a}{e^{2a_m}}
+a^2\sum_k A_k
\right)
\end{eqnarray}
and using the relations
\begin{eqnarray}
  \label{type A relatie 6}
A_i\beta_{im}^2+A_m\beta_{mi}^2-A_m\beta_{im}^2-A_i\beta_{mi}^2
&=& \frac{4}{e^{2a_i}}+\frac{4}{e^{2a_m}}
\\
A_m\beta_m+A_i\beta_i
&=&
8-4N + \frac{2N-2}{e^{2a_i}}+\frac{2N-2}{e^{2a_m}}
\label{type A relatie 7}
\end{eqnarray}
we find that the ninth and tenth lines are indeed exchanged under the interchange of $i$ and $m$. Therefore the prepotential (\ref{F A type}) satisfies the WDVV equations and we have proven half of theorem \ref{theorem A type}.

\subsubsection{The parameters $a= \frac{2}{N+1} $ , $b=-1$ , $c=N+1$}
In this section we prove the other half of theorem \ref{theorem A type}.
The third order derivatives of $F$ are given by
\begin{eqnarray}
  F_{klm}=\frac{2}{N+1} +\delta_{kl}\delta_{lm}K_k+\delta_{kl}\beta_{mk} +\delta_{km}\beta_{lk}+\delta_{lm}\beta_{kl}
\end{eqnarray}
with
\begin{eqnarray}
  K_k
&=&
\sum_{q \neq k}\beta_{kq}+\beta_k
\nonumber \\
\beta_k
&=&
\frac{-2}{1-e^{2a_k}}+2(N-1)
\nonumber \\
\beta_{ij}
&=&
\Biggl\{
\begin{array}{cc}
0 & \quad $if$ \quad i=j \\
2\frac{e^{2a_j}}{e^{2a_i}-e^{2a_j}} & \quad $if$ \quad i\neq j
\end{array}
\end{eqnarray}
Taking
\begin{eqnarray}
  h_j = e^{2a_j}+\sum_{i=1}^{N}e^{2a_i}
\end{eqnarray}
we find that up to a factor $B=\sum_j h_jF_j$ equals
\begin{eqnarray}
  \delta_{kl}\left( \frac{1}{1-e^{2a_k}} \right) =\delta_{kl}\frac{1}{A_k}
\end{eqnarray}
So with respect to the previous paragraph there are modifications in the definitions of $\beta_k$, $\beta_{ij}$ and $A_k$. This causes the relations (\ref{type A relatie 1}), (\ref{type A relatie 2}), (\ref{type A relatie 3}), (\ref{type A relatie 4}), (\ref{type A relatie 5}), (\ref{type A relatie 6}) and  (\ref{type A relatie 7}) to be changed to the following ones
\begin{eqnarray}
A_j\beta_{ij} +A_i\beta_{ji} &=& -2+2e^{2a_i}+2e^{2a_j}
\\
A_i \beta_{ij}+A_j\beta_{ji}&=&-2
\\
A_i\beta_{ji}\beta_{im} + A_j\beta_{ij}\beta_{jm}-A_m\beta_{jm}\beta_{im}&=&{4}{e^{2a_m}}
\\
A_i \beta_{ik}\beta_{im} + A_m\beta_{mk}\beta_{mi} + A_k\beta_{km}\beta_{ki}&=&4
\\
A_i\beta_{i}\beta_{im}+A_m\beta_m\beta_{mi} &=& 4N-8
\\
A_i\beta_{im}^2+A_m\beta_{mi}^2-A_m\beta_{im}^2-A_i\beta_{mi}^2
&=& {4}{e^{2a_i}}+{4}{e^{2a_m}}
\\
A_m\beta_m+A_i\beta_i
&=&
4N-8 + -(2N-2){e^{2a_i}}-(2N-2){e^{2a_m}}
\end{eqnarray}
and using these relations we find that the WDVV equations are again satisfied. This proves theorem \ref{theorem A type}.

\subsection{Other classical Lie algebras}
\label{section proof BCD type}
In this section theorem \ref{theorem BCD type} will be proven. Therefore we take a prepotential of the form
\begin{eqnarray}
  F(a_1,...,a_N) =\sum_{1\leq i<j \leq N} \biggl( f(a_i-a_j)+f(a_i+a_j) \biggl)
+\eta \sum_{i=1}^{N}f(a_i)
\end{eqnarray}
where again
\begin{eqnarray}
  f'''(x)=\coth(x)
\end{eqnarray}
The third order derivatives are given by (\ref{third order derivatives}) with
\begin{eqnarray}
    K_k
&=&
\sum_{q \neq k}\beta_{kq}+\beta_k
\nonumber \\
\beta_k
&=&
\eta \coth(a_i-a_j)
\nonumber \\
\beta_{ij}
&=&
\Biggl\{
\begin{array}{cc}
0 & \quad $if$ \quad i=j \\
\coth(a_i-a_j)+\coth(a_i+a_j) & \quad $if$ \quad i\neq j
\end{array}
\\
a&=&b=c=0
\end{eqnarray}
One can derive the following relations
\begin{eqnarray}
\label{BCD type relations 1}
  \beta_{ji}\beta_{im}+\beta_{ij}\beta_{jm}-\beta_{jm}\beta_{im}
&=&0
\\
\label{BCD type relations 2}
\beta_{ik}\beta_{im}+\beta_{mk}\beta_{mi}+\beta_{km}\beta_{ki}
&=&4
\\
\label{BCD type relations 3}
\beta_i\beta_{im}+\beta_m\beta_{mi} 
&=&2
\label{type BCD relaties}
\end{eqnarray}
which are identities that we will need later. Furthermore, we take
\begin{eqnarray}
  B_{kl}=\sum_{j=1}^{N} \sinh(2a_j)F_{jkl}
\end{eqnarray}
and using (\ref{BCD type relations 1}) we find
\begin{eqnarray}
  B_{kl} = \delta_{kl}\left( 1-N + \sum_{j=1}^{N}\cosh^2(a_j)
+\frac{1}{2}\left(2(N-2)+\eta \right)\cosh^2(a_k)  \right)
\end{eqnarray}
This becomes independent of $k$ and $l$ precisely for $\eta=-2(N-2)$. So for this value of $\eta$ we can regard $B$ as a multiple of the identity. First let us consider all other values of $\eta$, so that $B$ is equal to (\ref{general B}) with
\begin{eqnarray}
  A_k = \frac{1}{1-N + \sum_{j=1}^{N}\cosh^2(a_j)
+\frac{1}{2}\left(2(N-2)+\eta \right)\cosh^2(a_k)} = \frac{1}{X+Y_k}
\end{eqnarray}
In order to satisfy the WDVV equations, the expression (\ref{general expression}) should be symmetric in $i$ and $m$. Just as in the previous section, the first nontrivial condition is that the fifth and sixth lines of (\ref{general expression}) are exchanged under the interchange of $i$ and $m$. This condition translates into
\begin{eqnarray}
  A_i \beta_{ji}\beta_{im} + A_j\beta_{ij}\beta_{jm}-A_m\beta_{jm}\beta_{im}=0
\end{eqnarray}
and therefore
\begin{eqnarray}
  (X+Y_j)(X+Y_m)\beta_{ji}\beta_{im} + (X+Y_i)(X+Y_m)\beta_{ij}\beta_{jm}
-(X+Y_i)(X+Y_j)\beta_{jm}\beta_{im}=0
\end{eqnarray}
Working this out further we find
\begin{eqnarray}
  -\frac{1}{16}\frac{( e^{4a_i}-1)(e^{4a_j}-1)(2(N-2)+\eta)^2}{e^{2(a_i+a_j)}}=0
\end{eqnarray}
Therefore we find that for $\eta \neq -2(N-2)$ the WDVV equations are not satisfied. We will now determine what happens for the value $\eta=-2(N-2)$, for which $B$ becomes a multiple of the identity. Then (\ref{general expression}) becomes
\begin{eqnarray}
  \sum_{k=1}^{N}F_{ijk}F_{klm} 
&=&
\delta_{im}\biggl( \beta_{ji}\beta_{lm} + \delta_{ij}\beta_{lm}K_i
+\delta_{il}\beta_{ji}K_i + \delta_{ij}\delta_{lm} K_iK_m \biggl)
\nonumber \\
&+&
\delta_{jl}(1-\delta_{lm})\beta_{ij}\beta_{mj}
\nonumber \\
&+&
\delta_{jl}\delta_{lm}K_l\beta_{im}
\nonumber \\
&+&
\delta_{ij}\delta_{il}K_l\beta_{mi}
\nonumber \\
&+&
\delta_{il}(1-\delta_{jm})\beta_{ji}\beta_{mi}
\nonumber \\
&+&
\delta_{lm}(1-\delta_{ij})\left( \beta_{ji}\beta_{im}+\beta_{ij}\beta_{jm} \right)
\nonumber \\
&+&
\delta_{jm}(1-\delta_{il}) \beta_{ij} \beta_{lj}
\nonumber \\
&+&
\delta_{ij}(1-\delta_{lm})\left(  \beta_{li}\beta_{ml} + \beta_{mi}\beta_{lm} \right)
\nonumber \\
&+&
\delta_{ij}\delta_{lm} \biggl( K_i\beta_{im} +K_m\beta_{mi}
+\sum_{k\neq i,m}\beta_{km}\beta_{ki} 
\biggl)
\nonumber \\
&+&
\delta_{jm}\delta_{il} \left(\beta_{im}^2 +  \beta_{mi}^2 \right)
  \label{BCD type expression}
\end{eqnarray}
The seventh and eighth lines are exchanged under the interchange of $i$ and $m$ for the same reasons as the fifth and sixth lines.Therefore it remains to check that the ninth and tenth lines are exchanged. To do this, we use (\ref{BCD type relations 2}) and (\ref{BCD type relations 3}) and find
\begin{eqnarray}
  K_i \beta_{im}+K_m\beta_{mi}
&+&
\sum_{k \neq i,m}\beta_{km}\beta_{ki} \nonumber \\
&=&
\sum_{k \neq i,m}\left( \beta_{ik}\beta_{im}+\beta_{mk}\beta_{mi}+\beta_{km}\beta_{ki}  \right)
+\beta_{im}^2+\beta_{mi}^2+\eta (\beta_{i}\beta_{im}+\beta_{m}\beta_{mi})
\nonumber \\
&=&
\sum_{k\neq i,m} 4 +\beta_{im}^2+\beta_{mi}^2+2\eta = \beta_{im}^2+\beta_{mi}^2
+2\left( 2(N-2)+\eta \right)
\end{eqnarray}
So for the special value $\eta=-2(N-2)$ we can conclude that $F$ satisfies the generalized WDVV system. This finishes the proof of theorem \ref{theorem BCD type}.

\subsubsection{Adding a new variable}
\label{section proof adding}
We add a new variable to the prepotential of BCD type and consider the function
\begin{eqnarray}
\label{F adding proof}
F(a_1,...,a_{N+1})
&=&
\sum_{1\leq i<j \leq N} \biggl( f(a_i-a_j)+f(a_i+a_j) \biggl)
+\eta \sum_{i=1}^{N}f(a_i) 
\nonumber \\
&+&\gamma \left( 
\frac{1}{6}a_{N+1}^3+\frac{1}{2}a_{N+1}\sum_{k=1}^{N}a_K^2
\right)
\end{eqnarray}
The $B$ that we will use is
\begin{eqnarray}
  B_{kl}=\left(F_{N+1} \right)_{kl} = \gamma \delta_{kl}
\end{eqnarray}
and the WDVV condition becomes
\begin{eqnarray}
\label{adding condition}
  \sum_{k=1}^{N+1} F_{ijk}F_{klm} = \sum_{k=1}^{N+1} F_{mjk}F_{kli}
\qquad \qquad i,j,l,m=1,...,N+1
\end{eqnarray}
If $i=N+1$ or $m=N+1$ then this condition is automatically fulfilled. Restricting ourselves to $i,m \leq N$ we can rewrite the left hand side of (\ref{adding condition}) in the form
\begin{eqnarray}
  \sum_{k=1}^{N} F_{ijk}F_{klm} + F_{ij,N+1}F_{N+1,lm} =
  \sum_{k=1}^{N} F_{ijk}F_{klm} +\gamma^2 \delta_{ij}\gamma_{lm}
\end{eqnarray}
Therefore we can repeat the analysis of the previous paragraph, starting
with the expression (\ref{BCD type expression}). This expression is now changed by adding $\gamma^2$ to the $\delta_{ij}\delta_{lm}$ term in the penultimate line. Since this is the only change, the condition $\eta=-2(N-2)$ is modified to
\begin{eqnarray}
\label{parameters adding}
\label{modified condition}
  \eta=-2(N-2)-\frac{\gamma^2}{2}
\end{eqnarray}
This proves theorem \ref{theorem adding}.

\begin{remark}
The condition (\ref{parameters adding}) allows us to use the values $\eta=1,0$ which are associated with the $B,D$ Lie algebras. By adding the extra variable we can regard the corresponding prepotentials of the $B,D$ theories as solutions not only
to the generalized WDVV equations, but even to the original system (\ref{original WDVV}). Since $\eta$ is negative for real $\gamma$, the condition (\ref{modified condition}) requires $\gamma$ to be imaginary, and the prepotentials no longer
satisfy the property that real variables $a_k$ lead to a real value of $F$. To restore this property one can change the variables $a_k$ to $ia_k$ and change
$f$ to
\begin{eqnarray}
  f(x)= \frac{1}{6}(ix)^3 - \frac{1}{4}Li_3 (e^{-2ix}) = -\frac{i}{6}x^3 
-\frac{1}{4}\sum_{k=1}^{\infty} \frac{e^{-2ikx}}{k^3}
\end{eqnarray}
This $f$ is sometimes used in the literature, see e.g.
\cite{MIRO:2000}.
\end{remark}

\bibliographystyle{h-physrev}
\bibliography{5dimbiblio}
\include{thebibliography}

\end{document}